\documentclass[preprint,prl,showpacs,superscriptaddress]{revtex4-1}
\usepackage{amsmath}
\usepackage{graphicx}
\usepackage{amssymb}
\usepackage{times}
\usepackage{color}
\usepackage{multirow}
\definecolor{lr}{rgb}{1.0,0.3,0.3}
\definecolor{dg}{rgb}{0.0,0.5,0.0}

\usepackage{comment}

\graphicspath{./}
 


\begin{document}

\title{\emph{Ab initio} theory of negatively charged boron vacancy qubit in \emph{h}BN}

\author{Viktor Iv\'ady}
\thanks{Authors contributed equally}
\affiliation{Wigner Research Centre for Physics,  PO Box 49, H-1525, Budapest, Hungary}
\affiliation{Department of Physics, Chemistry and Biology, Link\"oping University, SE-581 83 Link\"oping, Sweden}

\author{Gergely Barcza}
\thanks{Authors contributed equally}
\affiliation{Wigner Research Centre for Physics, 
  PO Box 49, H-1525, Budapest, Hungary}
\affiliation{ J. Heyrovsk\'{y} Institute of Physical Chemistry, Academy of Sciences of the Czech Republic, Dolej\v{s}kova 3, 18223 Prague 8, Czech Republic}

\author{Gerg\H{o} Thiering}
\affiliation{Wigner Research Centre for Physics, 
  PO Box 49, H-1525, Budapest, Hungary}    

\author{Song Li}
\affiliation{Department of Mechanical Engineering, City University of Hong Kong, Kowloon Tong, Hong Kong, China}

\author{Hanen Hamdi}
\affiliation{Wigner Research Centre for Physics,
  PO Box 49, H-1525, Budapest, Hungary}
  
\author{\"{O}rs Legeza}
\affiliation{Wigner Research Centre for Physics, 
  PO Box 49, H-1525, Budapest, Hungary}

\author{Jyh-Pin Chou}
\affiliation{Department of Mechanical Engineering, City University of Hong Kong, Kowloon Tong, Hong Kong, China}

\author{Adam Gali} 
\email{gali.adam@wigner.mta.hu}
\affiliation{Wigner Research Centre for Physics,
  PO Box 49, H-1525, Budapest, Hungary}
\affiliation{Department of Atomic Physics, Budapest University of
  Technology and Economics, Budafoki \'ut 8., H-1111 Budapest,
  Hungary}

\date{\today}


\begin{abstract}
Highly correlated orbitals coupled with phonons in two-dimension are identified for paramagnetic and optically active boron vacancy in hexagonal boron nitride by first principles methods which are responsible for recently observed optically detected magnetic resonance signal. We report \textit{ab initio} analysis of the correlated electronic structure of this center by density matrix renormalization group and Kohn-Sham density functional theory methods. By establishing the nature of the bright and dark states as well as the position of the energy levels, we provide a complete description of the magneto-optical properties and corresponding radiative and non-radiative routes which are responsible for the optical spin polarization and spin dependent luminescence of the defect. Our findings pave the way toward advancing the identification and characterization of room temperature quantum bits in two-dimensional solids. 
\end{abstract}
\maketitle


\section*{Introduction}

Hexagonal boron nitrite ($h$BN) is a laminar van der Waals material with advanced fabrication techniques making it suitable for studying semiconductor physics in two dimension (2D). In particular, the wide energy gap of $h$BN may host numerous defects states with internal optical transitions that give rise to color centers. Some of these centers have already shown great potential in quantum technology application~\cite{Aharonovich2015,WrachtruphBN2016,Bassett2017,ProsciaOptica18,ExarhosNCOMMS2019,DyakonovArXiv2019,WrachtrupArXiv2019} demonstrated previously in the bulk semiconductors, such as diamond~\cite{DOHERTY2013,BecherPLR2018,FedorGeV2017,ST1NatNano} and silicon carbide~\cite{Gali10,Weber10,Koehl11,Soltamov12,Kraus2014}. Color centers in exfoliated $h$BN, however, may exhibit significant advances over their bulk counterparts owing to the distinct properties of quasi 2D semiconducting flakes. 

In recent years numerous color centers in $h$BN~\cite{Aharonovich2015, WrachtruphBN2016, FuchshBN2016, Bassett2017, ProsciaOptica18, BecherhBN2019, MFord2017, WestonPhysRevB2018, PlenioACS2018, Sajid2018} have been reported as new generation high temperature single photon emitters. While the microscopic configuration and electronic structure of these emitters are still not fully understood, it is widely accepted that these color centers can be associated with point defects and point defect complexes. Besides favorable optical properties, point defect often possesses a spin that can implement isolated quantum bits or qubits. So far previous investigations have focused on the optical properties of the color centers, the spin degree of freedom has been observed only in very recent experiments~\cite{ExarhosNCOMMS2019, DyakonovArXiv2019,  WrachtrupArXiv2019}. 

A key signature of potential point defect qubits is the optically detected magnetic resonance (ODMR) signal which makes optical spin initialization and read out possible through spin dependent decay processes from the excited state to the ground state. Recent experiments have recorded ODMR signal at room temperature for color centers in $h$BN~\cite{DyakonovArXiv2019, WrachtrupArXiv2019}. One of  the centers is associated with the negatively charged boron vacancy (VB) based on a previous theoretical study~\cite{PlenioACS2018}. Although, the charge transition levels~\cite{Bockstedte2011hBN, MFord2017, PlenioACS2018, WestonPhysRevB2018} and group theory analysis on the electronic structure aided by Kohn-Sham density functional theory (KS DFT) calculations~\cite{PlenioACS2018} have been reported for this defect but those studies neglected correlation effects and dynamic effects due to phonons and mostly focused on bright states. On the other hand, it has been recently shown from first principles for an exemplary point defect qubit, nitrogen-vacancy center in diamond (see Refs.~\onlinecite{ThieringPRB2018,BockstedteNPJ2018, Gali2019}) that phonon mixing of highly correlated dark states is a key in the optical spin polarization process. In addition, the spin-dependent properties of defect qubits, such as zero-field splitting (ZFS) and hyperfine constants (HF), have not yet been reported in $h$BN which is a key tool in identification of the point defect qubits and plays a crucial role in the quantum optical control of the qubit. 

In this study, we carry out a thorough first principles investigation on VB in single sheet \emph{h}BN. 
In particular, we compute zero-field splitting, hyperfine splitting, and photo luminescence spectrum, all of which are in good agreement with experiment. Many-body spectrum of VB coupled with phonons is demonstrated allowing us to discuss relevant spin selective decay processes that give rise to optical spin polarization and spin dependent luminescence of the defect. Our first principles results not only reproduce very recent experimental results on VB but also provide a theoretical foundation for developing VB based qubit applications.

\section*{Results}

The highest point group symmetry of VB in \emph{h}BN is  D$_{\text{3h}}$ that may reduce to C$_{2v}$ under the effect of external strain or due to Jahn-Teller distortion. The in-plane dangling bonds and the out of plane $p_z$ orbitals of the three neighbor nitrogen atoms give rise to a set of non-degenerate $a$ and degenerate $e$ single particle defect states. The six defect states are occupied with ten electrons in the negative charge state. Depending whether the defect state is formed by the in-plane dangling bonds or by the out of plane $p_z$ orbitals, we distinguish prime and double prime defect states, respectively. These prime and double prime states play similar role to the parity selection rule for systems with inversion symmetry in the optical transition.

\begin{figure}[h!]
\includegraphics[width=0.65\columnwidth]{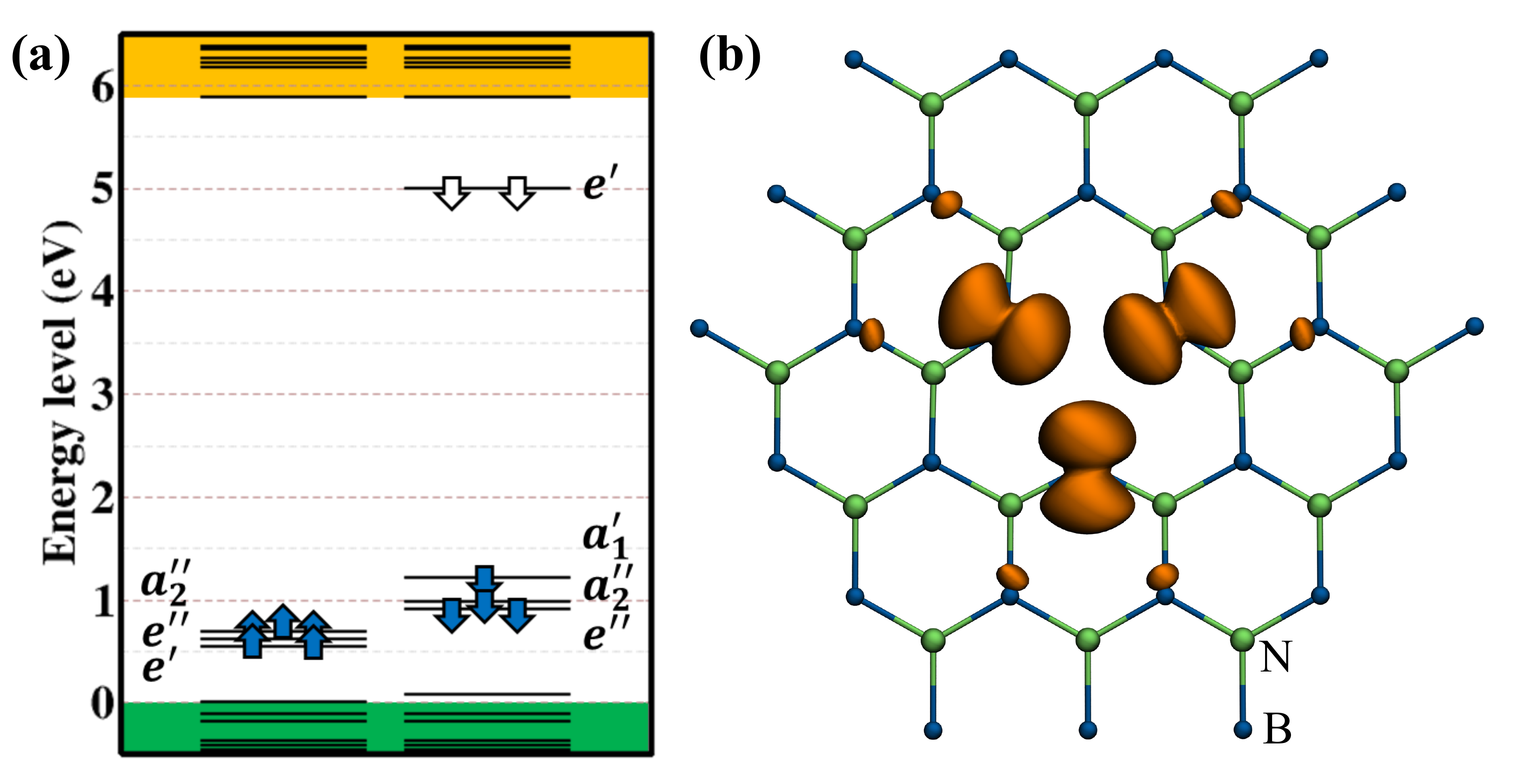}
\caption{ \label{fig:single} (a) Single particle electronic structure and (b) spin density of VB in \emph{h}BN.}
\end{figure}

Our first principles DFT calculation provides single particle Kohn-Sham (KS) electronic structure~\cite{PlenioACS2018} as depicted in Fig.~{\ref{fig:single}}(a). The electrons of VB occupy the defect states so that the $e'$ dangling bond state becomes half occupied and give rise to a spin-1 ground state. The spin density of the $^{3}{A_2}'$ ground state is depicted in Fig.~{\ref{fig:single}}(b), which shows similarities to the spin density of known qubits in 3D bulk materials such as the NV center in diamond and the divacancy in SiC~\cite{IvadyNPJComp}. Fine structure of the electron spin sublevels is dominated by the dipolar spin-spin zero-field interaction for which we obtained $D_s = +3.471$~GHz and $D_b = +3.467$~GHz in single sheet and bulk \emph{h}BN, respectively. These values are in remarkable agreement with the value of $ \left| D \right| = 3.4  $~GHz reported by recent experiments for the VB assigned color center in bulk $h$BN~\cite{DyakonovArXiv2019}. We note that the experimental results also show an $E = 40$~MHz splitting, which suggests the presence of strain in the sample or stray electric fields during the ODMR measurements. Furthermore, the sign of the $D$ constant was deduced to by negative in Ref.~[\onlinecite{DyakonovArXiv2019}] based on indirect measurements. Our results question the validity of findings. Nuclear spins of the host $h$BN lattice give rise to hypefine splitting of the spin sublevels. In Table~S1-S2, we provide the calculated hyperfine coupling constants of nitrogen and boron nuclear spins for the most relevant neighboring sites of VB, see Fig.~S1. In the calculations, we obtain $A_{z,s} = 47.9$~MHz and $A_{z,b} = 47.2$~MHz splitting for the first neighbor $^{14}$N nuclei in single sheet and bulk $h$BN, which are in excellent agreement with the experimental observation of 47~MHz~\cite{DyakonovArXiv2019}. Our \emph{ab initio} ground state spin coupling parameters thus unambiguously identify the experimental color center reported in Ref.~[\onlinecite{DyakonovArXiv2019}] as VB. Furthermore, our results demonstrate that the ground state coupling parameters are essentially identical in free standing single sheet and bulk $h$BN.

\begin{figure}[h!]
\includegraphics[width=0.95\columnwidth]{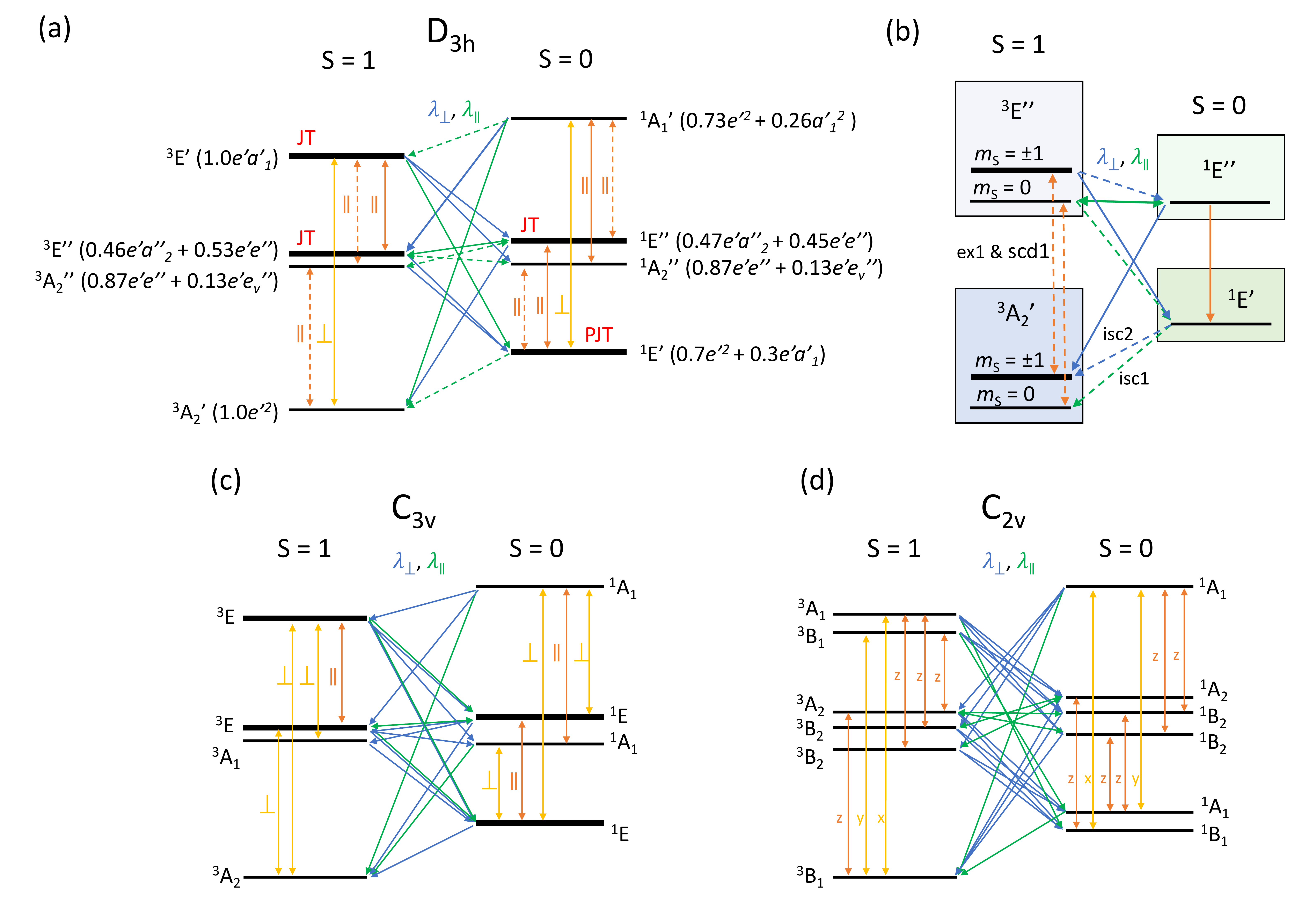}
\caption{ \label{fig:MBels} Electron structure and decay pathways of VB. (a) Many-body electron structure as obtained on the D$_{3h}$ symmetric ground state configuration. Orange and red arrows indicate perpendicular and parallel polarized optical transitions, while blue and green arrows indicate coupling between singlet and triplet states via perpendicular and parallel spin-orbit coupling, respectively. In all cases, leading order Slater determinant expression of the electron hole density is provided. JT and PJT labels Jahn-Teller(JT) and partial Jahn-Teller (PJT) states. Dashed arrows highlight transitions allowed by JT effects. (b) Simplified electronic structure for understanding spin dependent non-radiative decay processes. Dashed arrows indicate second order, JT, PJT, and strain allowed transitions. ex1 and scd1 label optical excitation and spin conserving decay, while isc1 and isc2 label inter system crossings relevant for the optical spin polarization, respectively. (c)-(d) Many-body electron structure as generalized to strained cases by considering (c) C$_{3v}$ symmetric and (d) C$_{2v}$ symmetric reconstructions.}
\end{figure}

To obtain the many-body spectrum and the corresponding excited states, we apply the inherently multireference density matrix renormalization group method (DMRG)~\cite{White-1992b} generalized for arbitrary long-range interactions together with optimization tools based on quantum information theory \cite{Legeza-2003b} using Kohn Sham orbitals to span the active space. For further details, see Methods. The vertical excitation spectrum as obtained on the ground state configuration of VB is provided in Table~S3 and depicted in Fig.~\ref{fig:MBels}(a). In the same figure, we provide leading order expression of the excited states electron hole density.  We note that the spectrum shows similarities but also difference to previous reports which did not account for high correlation effects~\cite{PlenioACS2018}. In the triplet manifold, we find two excitation branches. The lower branch contains nearly degenerate $^{3}A_2^{\prime\prime}$ and $^{3}E^{\prime\prime}$ excited states that accommodate the hole of the excited electron on the $e^{\prime\prime}$ state of the defect mixed with the valence band $e^{\prime\prime}$ states and on the mixture of $e^{\prime\prime}$ and $a_2^{\prime\prime}$ defect states, respectively. The higher lying branch contains a single state $^{3}E^\prime$ which account for excitation from the $a_1^{\prime}$ state to the $e^{\prime}$ state of VB. In the singlet manifold, we find $^{1}E^\prime$ state with the lowest energy, however, the $^{1}A_1^\prime$ state of similar determinant appears at very high energy. In-between these states we find the singlet counterparts of the $^{3}A_2^{\prime\prime}$ and $^{3}E^{\prime\prime}$ excited states, i.e.\  $^{1}A_2^{\prime\prime}$ and $^{1}E^{\prime\prime}$ excited states. These results demonstrate that static correlation effects play crucial role in forming the excited state spectrum. We note that orbital and structural relaxation effects as well as dynamical screening effects decrease the many-body vertical excitation energies\cite{BockstedteNPJ2018} provided in Table~S3. 

Based on group theory considerations in $D_{3h}$ symmetry of the ground state, we find several allowed optical transitions, yellow and orange arrows in Fig.~\ref{fig:MBels}(a), and spin-orbit interactions, blue and green arrows in Fig.~\ref{fig:MBels}(a), that may allow spin dependent nonradiative transitions. In the following, we restrict our considerations to the lowest three excited states, $^{3}{A_{2}}^{\prime\prime}$, $^{3}E^{\prime\prime}$, and $^{3}E^{\prime}$, in the triplet manifold and to the three lowest  excited states, $^{1}E^\prime$, $^{1}A_{2}^{\prime\prime}$, and $^{1}E^\prime$, in the singlet manifold. 

First, we investigate optical decay processes that may account for the photoluminescence spectum reported recently for this center~\cite{DyakonovArXiv2019}. Note that the lowest energy excited states in the triplet manifold, $^{3}{A_2}^{\prime\prime}$ and $^{3}{E}^{\prime\prime}$, are optically not allowed~\cite{PlenioACS2018} and transition from the $^{3}A_2^\prime$ ground state is only possible to the higher lying $^{3}E^\prime$ states with perpendicular polarization. These selection rules can be relaxed by strain and electric field that may reduce the symmetry to C$_{3v}$ and C$_{2v}$ depending on the character of the external field and thus give rise to additional optical excitation pathways, see Fig.~\ref{fig:MBels}(c)-(d). Accordingly, $^{3}E^{\prime\prime}$ and $^{3}A_2^{\prime}$ states transform to $^{3}E$ and $^{3}A_2$ in C$_{3v}$ allowing perpendicular polarized photon emission and absorption, while the $^{3}E^{\prime\prime}$ state splits as $^{3}A_2$ and $^{3}B_2$ in C$_{2v}$ allowing parallel polarized photon mediated photon transition between the $^{3}B_1$ ground state and the $^{3}A_2$ excited state. 

Similarly to the effect of strain,  $e^\prime \otimes E^{\prime\prime}$  Jahn-Teller (JT) effect may give rise to additional transitions. In Fig.~\ref{fig:MBels}(a), we marked all the JT unstable states in D$_{3h}$ symmetry. Allowed transitions due the coupling with $e^\prime$ phonon modes are marked by dashed arrows in Fig.~\ref{fig:MBels}(a). The dark $^{3}E^{\prime\prime}$ state combined with an $e^\prime$ local vibration mode may give rises to polaronic states that can be connected with the ground state through parallel polarized photon emission and absorption.

\begin{table}
\begin{ruledtabular}
\caption{\label{tab:ESdata} Low energy spectrum as obtained by HSE06 methods after structural optimization. All values are in eV. The spectrum is referenced to the ground state. We note that the  $^{3}E^{\prime}$ state the inaccuracy in the calculated excitation energy might be larger than for the lower energy states (see Methods).}
 \begin{tabular}{c|ccc }
   &  D$_{3h}$ & C$_{2v}$ & C$_{1}$ \\ \hline
  $^{3}A_2^\prime$ & 0.00 &  &  \\
  $^{1}E^\prime$  &  0.515  & 0.383 & 0.383 \\
  $^{3}E^{\prime\prime}$  &  1.916 &  1.723  & 1.710 \\
  $^{1}E^{\prime\prime}$  &   2.005 & 1.708 & 1.701 \\
  $^{3}E^{\prime}$ &   2.287 & &  \\
 \end{tabular}
\end{ruledtabular}
\end{table}

Structural relaxation and JT effects in the excited states can be investigated by \emph{ab initio} DFT methods, for details see Methods. Indeed, our DFT simulations reveal considerable JT distortions that release 0.193~eV, 0.297~eV, and 0.132~eV energy in the $^{3}E^{\prime\prime}$, $^{1}E^{\prime\prime}$, and $^{1}E^\prime$ states, see Table~\ref{tab:ESdata}. We note that the $^{1}E^\prime$ singlet state, dominated by the ${e^\prime}^2$ determinant, is JT unstable only due to the inter mixture of JT unstable  $e^\prime a_1^\prime$ Slater determinant. 
Besides $e^\prime$ phonons, membrane vibrational modes\cite{GichangNanoLett2018} also couple to the electronic structure of VB in the $^{3}E^{\prime\prime}$ and $^{1}E^{\prime\prime}$ states that further reduce the symmetry and the energy, see 
Table~\ref{tab:ESdata}. While the JT and membrane distortions of the first neighbor shell of VB are in the same order, $\approx 0.1$~\AA, the energy gain due to out of plane distortions is an order of magnitude smaller than the JT energy. This indicates that the potential energy landscape is considerably flatter for out of plane structural relaxations. The $^{1}E^{\prime}$ state partially inherits this behavior by mixing with $^{1}E^{\prime\prime}$ state when planar reflection symmetry is broken. The resultant out of plane relaxation is $\approx 0.01$~\AA\   for the first neighbor nitrogen atoms, while the energy gain is smaller than 1 meV.

\begin{figure}[h!]
\includegraphics[width=0.65\columnwidth]{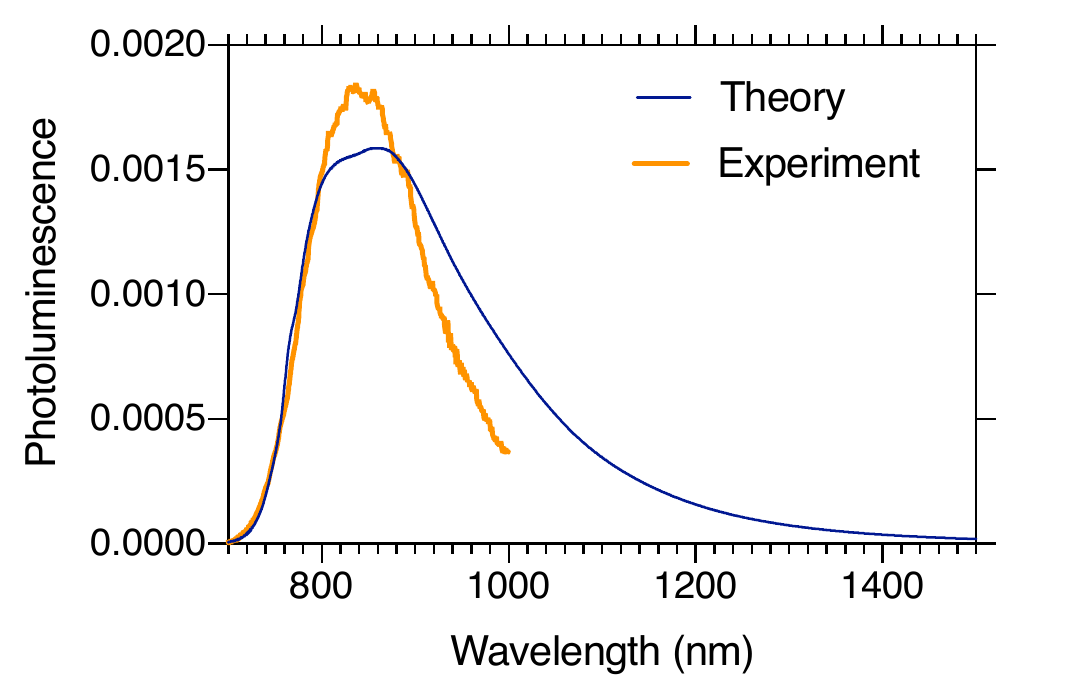}
\caption{ \label{fig:PL}  Photoluminescence spectrum of VB at 300~K as obtained from \emph{ab initio} simulations for the $^{3}E^{\prime\prime} \rightarrow {^{3}A_2^\prime}$ transition. To produce the theoretical curve the calculated Huang-Rhys factor of 3.5 is used whereas the position of ZPL at 765~nm (1.62~eV) was aligned to match to the highest intensities of the experimental spectrum which corresponds to about 0.09~eV redshift compared to the KS DFT result (1.71~eV). This difference is within the usual inaccuracy of the KS DFT method. The experimental spectrum is reported in Ref.~\onlinecite{DyakonovArXiv2019}.}
\end{figure}

Due to the fact that $^{3}A_2^{\prime\prime}$ and $^{1}A_2^{\prime\prime}$ states do not experience JT distortion, whereas  $^{3}E^{\prime\prime}$ and $^{1}E^{\prime\prime}$ states relax substantially due to JT effect, we anticipate that the latter state are lower in energy after geometry reconstruction. The spectrum of VB as obtained on corresponding relaxed excited state configurations is provided in Table~\ref{tab:ESdata}. Indeed, $^{3}E^{\prime\prime}$ and $^{1}E^{\prime\prime}$ are the first and second excited states in the triplet and singlet manifold, respectively. We find allowed optical transitions with parallel polarization and zero phonon luminescence (ZPL) energies 1.710~eV and 1.318~eV between these states and the ground and the lowest energy singlet state. The calculated 300~K emission side band of the optical transition in the triplet manifold as approximately determined by ground state phonon modes is depicted in Fig.~\ref{fig:PL}(a). The spectrum of Huang-Rhys factor at 3.5 fairly agrees with the experimental spectrum. Disagreement between theory and experiment may be related to lower detection efficiency at high wavelength ($>1000$~nm) in the experiment. We obtain best agreement between the spectra by assuming 1.62~eV energy (765~nm) for the ZPL emission, which is not resolved in the experiment. This value and the first principles triplet ZPL excitation energy agree very well, thus we assign the experimental optical spectrum to the optical transition from the $^{3}E^{\prime\prime}$ JT system to the  $^{3}A_2^\prime$ ground state. 

To study the polaronic solution of the triplet excited state in more details, we first calculated the full adiabatic potential energy surface and we obtained barrier energy between different JT distortions of 152~meV. By solving the $e^\prime \otimes E^{\prime\prime}$ JT Hamiltonian~\cite{Thiering2017}, we obtained the polaronic spectrum as listed in Table~S4. The first polaronic $\widetilde{A_1^{\prime\prime}}$ excited state lies only 0.01~$\mu$eV above the polaronic $\widetilde{E^{\prime\prime}}$ ground state that indicates effectively static JT distortion of the $^{3}$E$^{\prime\prime}$ state. The $\widetilde{A_1^{\prime\prime}}$ state has an allowed optical transition towards the electronic $A_2^{\prime}$ ground state by emitting a photon with parallel polarization. This finding justifies the application of Huang-Rhys theory for calculating the ZPL energy and the phonon sideband of the photo luminesence (PL) spectrum as explained in Ref.~\onlinecite{Gali2019}. The calculated radiative lifetime is about 20~$\mu$s in the distorted structure that we obtained in single particle approximation. This finding is in accordance to the second order optical transition. We note that the calculated absorption from the electronic ground state is about 3 orders of magnitude larger for the optically allowed $^{3}E^{\prime}$ state. Thus, optical driving of the center can be much more effective with photoexcitation towards the second triplet excited state $^{3}E^{\prime}$ and then the electron will decay to the lower $^{3}E^{\prime\prime}$ state. Indeed, the 2.33~eV excitation used in the experiment~\cite{DyakonovArXiv2019} is sufficient to excite the system to the optically allowed $^{3}E^{\prime}$ state, see Table~\ref{tab:ESdata}.

The excited state structure of VB offers several alternative decay pathways from triplet optically excited states through singlet excited states to the ground state. We describe the most relevant ones by a six-state rate equation model in supplementary section~IV. The spin selectivity of the decay is predominated by the strong coupling between the triplet and singlet $E^{\prime\prime}$ excited states that we find nearly degenerate, have akin atomic configurations, and couple with parallel spin-orbit interaction of approximately 170~GHz strength as obtained from KS DFT calculation. This corresponds to several orders of magnitude larger rate than the calculated radiative rate between the triplets. Due to this strong coupling and the first order allowed optical transition between the lowest energy singlets, we infer that the decay through the red-shifted singlet optical emission, with ZPL at $\approx$1.32~eV (939~nm), dominates over the triplet optical decay for the $m_\text{S} = 0 $ spin sublevel. The sign of the ODMR contrast predominantly determined by the slow second order decay processes, such as the direct optical and non-radiative $^{3}E^{\prime\prime} \rightarrow {^{3}A}_2^{\prime}$ transition with rate $r_{scd1}$ and the $^{1}E^{\prime} \rightarrow {^{3}A}_2^{\prime} \left( m_{\text{S} = 0} \right)$ transition with rate $r_{isc1}$, see supplementary section~IV. The negative contrast reported in the experiment\cite{DyakonovArXiv2019} is in agreement with the expectation  that $r_{isc1} < r_{scd1}$. 

Optical pumping induced ground state spin polarization is governed by the $^{1}E^{\prime} \rightarrow {^{3}A}_2^{\prime} \left( m_{\text{S}} = \pm 1 \right)$ transition rate $r_{isc2}$ and $r_{isc1}$. We note that these rates correspond to such decay processes that are allowed only in second order due to out of plane relaxation and partial JT distortion, respectively. This characteristic infers that the result of optical pumping can depend very much on external conditions, such as strain and electric field, that may influence rates $r_{isc1}$ and $r_{isc2}$.  We can distinguish two different scenarios. When $r_{isc2} > c r_{isc1}$, where $c$ is a prefactor between 0 and 1 determined by rate $r_{scd1}$ and  rate $r_{isc4}$ of ${^{3}E}^{\prime\prime} \left( m_{\text{S}} = \pm 1 \right) \rightarrow {^{1}E^{\prime}}$,  the spin is polarized in the $m_{\text{S}} = \pm 1 $ sublevels, for which we expect luminescence in the triplet manifold with ZPL of $\approx$1.71~eV energy. For $c r_{isc1} > r_{isc2}$, the spin is polarized in the $m_{\text{S}} = 0 $ sublevel of the ground state for which we expect red-shifted luminescence through the singlet states. As the reported room temperature luminescence fits to the triplet transition, the former scenario is more probable and the spin should be polarized in the $m_{\text{S}} = \pm 1 $ sublevels of the ground state accordingly. This finding is supported by the reported out of plane relaxation of the first neighbor atoms at equilibrium and the soft potential for membrane vibrational modes and out of plane distortions.

The above described spin selective decay mechanisms demonstrate  distinct behavior of VB from the known solid state qubits. The features provided by VB can be utilized, for example, to boost the ODMR contrast and to control the optical spin polarization mechanism. For achieving enhanced ODMR contrast one may use conventional optical filters to separate triplet emission associated with $m_{\text{S}} = \pm 1$ states and the red shifted singlet emission associated with $m_{\text{S}} = 0$ state. This way the ODMR contrast may be significantly increased. Furthermore, additional excitation in the singlet manifold can enhance photon count rate for the $m_{\text{S}} = 0$ sublevel to speed up the read out process. As the ground state spin polarization is determined by second order transitions induced by structural distortions, application of strain may allow control over the spin polarization and thus the photo luminescence of the center. For instance, strain that breaks planar reflection symmetry may enhance $m_{\text{S}} = \pm 1$ polarization, while strain of $E^{\prime}$ symmetry may result in $m_{\text{S}} = 0$ polarization and red-shifted luminescence.

\section{Discussion}

Our \emph{ab initio} results explain the known magneto-optical properties of VB in $h$BN. These achievements validate the electronic structure and spin polarization mechanism described in the results section. Having this in hand, one may apply more advanced techniques to harvest functionalities allowed by the electronic structure of VB. So far, only one optical transition is observed in the experiment, however, we predict multitude of them. Application of two or more excitation lasers may be utilized to control and improve optical initialization and read out properties of the defect. Furthermore, we predict extraordinaire capability for VB to control its magneto-optical properties by engineering strain.

We note that our calculations were performed dominantly in a single sheet model, while the experiment were carried out in bulk samples. The good agreement between theory and experiment indicates that VB should function in free standing single sheet $h$BN as good as in bulk, which is an important requirement for low dimension applications.

Finally, in our study, we utilized a so far less recognized combination of methods: density matrix renormalization group method on active space spanned by localized defect orbitals defined by hybrid density functional calculations.  Successful application of the method on the involved electronic structure of VB defect demonstrate that this method may be a key tool in investigating functional color centers in $h$BN and in other wide band gap semiconductors.

\section*{Methods}

We apply two methods, density functional theory (DFT) and density matrix renormalization group (DMRG), to describe the electronic structure of VB on different levels of theory. For the DFT calculations plane wave basis set of 450~eV and PAW~\cite{PAW} atomic potentials are used as implemented in VASP~\cite{VASP} as well as plane wave basis set of 750~eV and norm-conserving pseudo potentials are used as implemented in Quantum Espresso~\cite{QE-2017}. HSE06 hybrid functional~\cite{HSE03} with 0.32 exact exchange fraction~\cite{WestonPhysRevB2018} is used for hyperfine calculations~\cite{Szasz13}, excited state calculation in the framework of constrained occupation DFT~\cite{Gali:PRL2009}, and structural relaxation. We use 162 atom super cell of single sheet $h$BN embedding a single boron vacancy. In perpendicular direction, we use 30~\AA\ supercell size. Our bulk $h$BN model consists of 972 atoms ($9 \times 9 \times 3$ primitive cells) and includes a single boron vacancy.  HSE06 functional is used to calculate hyperfine\cite{Szasz13} and spin-spin zero-field-splitting parameters. To eliminate spin contamination in the latter case, we apply the correction scheme proposed in Ref.~[\onlinecite{BiktagirovArxiv}]. To determine the phonon spectrum of the ground state VB configuration we use PBE functional. Phonon sideband and polaron spectrum are obtained by the machinery described in Ref.~\onlinecite{ThieringPRB2018}. 

The studied sample, consisting of more than a hundred atoms, is described by several hundreds of Kohn-Sham orbitals making prohibitively expensive to be tackled directly by the DMRG approach. Therefore, a careful selection of active space to be treated on post-DFT level is important describing the remaining single particle states, i.e., which are not included in the active space, in frozen core approximation.
Based on previous studies on defect systems\cite{BockstedteNPJ2018}, it is expected that the low lying energy spectrum is predominantly determined by the highly localized defect orbitals. Therefore, the discussed results are based on DMRG calculations performed on active space of 27 spin restricted Kohn-Sham DFT orbitals which is selected according to the localization of the orbitals around the defect, i.e., occupied orbitals kept in the active space whose degree of localization on the first neighbor three nitrogen atoms reaches at least 0.10. In order to preserve the point group symmetry of the orbitals, we studied canonical orbital set without any further localization.

To verify the applicability of the active space selection based on localization, we also performed DMRG calculations defining the active orbitals based on an energy window around the localized defect levels. The structure of the many-body excited state spectrum yielded on such active space is essentially identical to the presented one observing upward shifts in energy in the range of 0.01-0.15~eV.  Obtaining slightly lower energies with the applied variational computational procedure, e.g., by $-$0.06~eV for the ground state, indicates that the orbital selection based on localization is preferable to the one based on energy.

One particle and two particle integrals are calculated by our in-house code uses Quantum Espresso Kohn-Sham orbitals obtained by spin restricted PBE calculations on fixed geometries provided by VASP HSE06 calculations. DMRG calculations are performed using the Budapest DMRG code~\cite{budapest_qcdmrg}.
Considering the large number of excited states to be computed, in order to enhance convergence, distinct DMRG calculations are performed fixing the total spin of the target states to 0 and 1. The numerical accuracy of the calculations was controlled by using elements of quantum information theory and by the dynamic block state selection approach~\cite{Szalay-2015a} keeping up to thousands of block states for the priory set quantum information loss threshold value $\chi=10^{-5}$. The density matrix of the system block in the DMRG truncation procedure is formed of the equally weighted linear combination of all target states.

In the hybrid DFT constrained occupation calculations proper excited state configurations are identified by comparing spin density and partial charge density provided by DFT calculations with the true many body spin and charge densities obtained from DMRG calculations. We note that constrained occupation hybrid DFT calculations often provide non-physical solutions. In particular, this is the case for higher lying excited states. Furthermore, the triplet and singlet $E^{\prime\prime}$ states are multi-determinant as provided by DMRG calculation. We approximate this state by setting the excited hole as a mixture of $a_2^{\prime\prime}$ and $e^{\prime\prime}$ states. Note that the spin density of this mixture is a good approximation to the spin density of the many-body excited state.

\section*{Acknowledgments} 

VI acknowledges the support from the MTA Premium Postdoctoral Research Program. This work was financially supported by the Knut and Alice Wallenberg Foundation through WBSQD2 project (Grant No.\ 2018.0071). GB is supported by NKFIH PD-17-125261 and “Bolyai” Research Scholarship of HAS. \"OL acknowledges grant NKFIH K120569 and the support of the Alexander von Humboldt Foundation. AG acknowledges the Hungarian NKFIH grants No.\ KKP129866 of the National Excellence Program of Quantum-coherent materials project. \"OL and AG acknowledge support of the NKFIH through the National Quantum Technology Program (Grant No. 2017-1.2.1-NKP-2017-00001). The calculations were performed on resources provided by the Swedish National Infrastructure for Computing (SNIC 2018/3-625 and SNIC 2019/1-11) at the National Supercomputer Centre (NSC) and by the Wigner RCP.


%

\end{document}